# Complexes of DNA Bases and Gold Clusters Au$_3$ and Au$_4$ Involving Nonconventional N-H···Au Hydrogen Bonding

E. S. Kryachko[*,†] and F. Remacle[*,‡]

Department of Chemistry, B6c, University of Liège, B-4000 Liège, Belgium

and Bogoliubov Institute for Theoretical Physics, Kiev, 03143 Ukraine

## Abstract

DNA base-gold interactions are studied theoretically at the DFT level using Au$_3$ and Au$_4$ clusters as simple catalytic models for Au particles. The bonding between DNA bases and gold clusters occurs via the anchoring of a Au atom to the N or O atoms of the bases. In the most stable planar base-Au$_3$ complexes, the Au-N or Au-O anchor bonds are reinforced by N-H···Au bonds. The mechanism of formation of these nonconventional H-bonds is discussed.

---

[*] Corresponding author.
[†] University of Liège and Bogoliubov Institute for Theoretical Physics;
  E-mail: eugene.kryachko@ulg.ac.be

[‡] Maître de Recherches, FNRS (Belgium). University of Liège;
  E-mail: fremacle@ulg.ac.be

DNA - gold interactions play an important role in nano- and biotechnology (see[1-5] and references therein). Recent experimental studies have shown that the DNA bases, adenine (A), thymine (T), guanine (G), and cytosine (C) interact with Au surfaces and Au clusters, and adsorb at Au electrodes in a complex and sequence-dependent manner.[4,5] Their relative affinities to adsorb on polycrystalline Au films are ordered as A > C ≥ G > T.[4a] Mirkin and co-workers[4b] have recently reported that the heats of desorption of the nucleic acid bases from Au thin films vary within 26.3 - 34.9 kcal·mol$^{-1}$ and obey the following order: G > A ≥ C > T. Chen et al. [4d] and Giese and McNaughton [4e] have suggested that adenine binds Au via the $N_6$ exocyclic amino group and the $N_7$ atom. However, the mechanism of binding of DNA bases with gold as well as the precise geometries of their complexes remain still unknown. These issues are addressed in the present work.

We investigate theoretically the interaction of DNA bases with bare $Au_3$ and $Au_4$ gold clusters as simple catalytic models of Au particles[6], using the density functional B3LYP potential together with the energy-consistent 19-valence electron relativistic effective core potential for gold developed by Ermler, Christiansen, and co-workers[7] (see also[8] for current work and references therein) and the 6-31+G(d) basis set for DNA bases, as they are implemented in the GAUSSIAN 03 package of quantum chemical programs.[9] All geometrical optimizations use the keywords "tight" and "Int=UltraFine". The harmonic vibrational frequencies, zero-point vibrational energies (ZPVE), and enthalpies were also calculated. The reported binding energies include the ZPVE correction. In order to assess basis-set effects, some selected DNA base-gold structures were further studied at the B3LYP/RECP (gold) ∪ 6-31++G(d,p) (DNA base) computational level.

Single gold atom – DNA bases complexes are characterized by very low binding energies: adenine (2.2 - 2.5 kcal·mol$^{-1}$), guanine (0.8 kcal·mol$^{-1}$), and cytosine (1.2 and 3.8 kcal·mol$^{-1}$), whereas no stable complex is formed with thymine. On the other hand, the binding energy of a bare $Au_3$ gold cluster to DNA bases (see Table 1) is about an order of magnitude higher. The most stable base-$Au_3$ complexes are planar and formed either via Au-N or Au-O coordinative ("anchor") bonds (Figures 1-4). The $Au_3$ cluster preferentially binds adenine at $N_1$, $N_3$, and $N_7$ (figure 1), thymine at $O_2$ and $O_4$ (figure 2), guanine at the $N_3$, $O_6$ and $N_7$ atoms (figure 3), and cytosine at $O_2$ and $N_3$ (figure 4). The binding energies are larger (they fall within the interval of 20.7 - 25.4 kcal·mol$^{-1}$), for complexes with a Au-N anchor bond than for complexes with a Au-O one (for the latter, $E_b \in$ 10.5 - 20.0 kcal·mol$^{-1}$), indicating that the Au-N bonding is stronger. Correspondingly, the Au-N bonds are also shorter, ranging from 2.130 to 2.153 Å, compared to the Au-O ones (2.177 - 2.239 Å). The most stable complex has a binding energy $E_b$ = 25.4 kcal·mol$^{-1}$ and is formed between cytosine and $Au_3$ anchored to the $N_3$ atom on the $N_2$ side. The complexes A·$Au_3$, G·$Au_3$, and C·$Au_3$ with the $Au_3$ - amino nitrogen bonding (also shown in Figures 1, 3, and 4) are nonplanar and less stable, viz., $E_b$(A·$Au_3$($N_6$)) = 14.5, $E_b$(G·$Au_3$($N_2$)) = 9.1, and $E_b$(C·$Au_3$($N_4$)) = 11.2 kcal·mol$^{-1}$. In agreement with the experiments by Lindsay and co-workers[1a] (see also[4g]), the complexes between T and $Au_3$ anchored to the $N_1$ and $N_3$ atoms are unstable. A comparison of the binding energies of planar and nonplanar base-$Au_3$ complexes leads to the following order for the affinities of the DNA bases to $Au_3$: C > A > G > T. Since the purines A and G possess four and six anchoring sites for $Au_3$, respectively, and the pyrimidines C and T have only three, the mean values of the binding energies, averaged over the number of sites, are reordered as A > C > G > T.

The binding energies of the most stable base-$Au_3$ complexes and their anchor bond lengths are not fully correlated (see Table 1). For example, the most stable complex C·$Au_3$($N_3$) shown in Figure 4 has a Au-N anchor bond of 2.164 Å whereas the shortest one, 2.130 Å, is found in A·$Au_3$($N_7$) whose $E_b$ = 22.3 kcal·mol$^{-1}$. The reason is that the anchor bonds in all reported planar base-Au$_3$ complexes, except G·Au$_3$(N$_7$), are reinforced by the N-H···Au bonds. Note that a similar bonding may also occur between the amino groups and Au$_3$ in the complexes A·Au$_3$(N$_6$), G·Au$_3$(N$_2$), and C·Au$_3$(N$_4$) (see Figures 1, 3 and 4) but these bonds are much weaker since their corresponding bond angles ∠NHAu ≈ 100°. Regarding the aforementioned complex G·Au$_3$(N$_7$), an extremely weak, C$_8$-H$_8$···Au$_{11}$, bond may also be formed therein due to the very weak proton donor ability of the C$_8$-H$_8$ group[10e].

The non conventional N-H···Au bonds that are present in the most stable complexes are similar to conventional weak hydrogen bonds since they obey all the necessary prerequisites (see[10] and references therein) of the conventional ones, viz.: (i) there exists an evidence of the bond formation; (ii) there exists an evidence that this bond specifically involves a hydrogen atom which is bonded to Au predominantly along the N-H bond direction; (iii) the N-H bond elongates relative to that in the monomer; (iv) there exists a van der Waals cutoff: the hydrogen bond separation r(H···Au) is shorter than the sum of van der Waals radii of H and Au (2.86 Å); (v) the stretching mode ν(N-H) undergoes a red shift with respect to that in the isolated base and its IR intensity increases; and finally, (vi) proton nuclear magnetic resonance ($^1$H NMR) chemical shifts in the N-H···Au hydrogen bond are shifted downfield compared to the monomer.[11] The key features of the N-H···Au bonds of the planar base-Au$_3$ complexes that are gathered in Table 1 demonstrate that these bonds satisfy all the conditions (i) - (vi). Therefore, they can be treated as the nonconventional hydrogen ones, by analogy with other nonconventional hydrogen bonds with transition metals.[12]

Let us now focus on some specific features of these bonds. Among all the studied complexes A·Au$_3$, the largest red shift of the ν(N$_9$-H$_9$) stretch (252 cm$^{-1}$) is predicted for the hydrogen bond N$_9$-H$_9$···Au$_{11}$ in A·Au$_3$(N$_3$) (see Figure 1). This can be understood from the fact that the N$_9$-H$_9$ bond of A is characterized by the smallest deprotonation energy (enthalpy) compared to the other N-H bonds, viz., DPE($N_9$-$H_9$) = 336.8 kcal·mol$^{-1}$ < DPE($N_6$-$H_6'$) = 355.2 kcal·mol$^{-1}$ < DPE($N_6$-$H_6$) = 355.8 kcal·mol$^{-1}$,[13] and, therefore, it is more strongly perturbed by the $Au_3$ cluster than the amino protons. The three shortest N-H···Au H-bonds (2.580 – 2.627 Å) are found in the complexes G·$Au_3$($O_6$; $N_1$ side), T·$Au_3$($O_2$; $N_1$ side), and C·$Au_3$($O_2$; $N_1$ side). The corresponding N-H stretching modes are characterized by the largest red shifts (302, 324, and 306 cm$^{-1}$), the largest IR enhancement factors (15, 11 and 14), and the largest displacements, 0.014 - 0.017 Å, of the bridging proton toward the gold atom acting as a nonconventional proton acceptor. The high stabilization of these hydrogen bonds is explained by the small deprotonation energies of the involved N-H groups: DPE($N_1$-$H_1$; G) = 338.4 kcal·mol$^{-1}$, DPE($N_1$-$H_1$; T) = 334.2 kcal·mol$^{-1}$ < DPE($N_3$-$H_3$; T) = 346.6 kcal·mol$^{-1}$, and DPE($N_1$-$H_1$; C) = 345.2 kcal·mol$^{-1}$ < DPE($N_4$-$H_4$; C) = 354.2 kcal·mol$^{-1}$.[13] The former inequality also partly explains the difference of 3.6 kcal·mol$^{-1}$ in the binding energies of the complexes T·$Au_3$($O_2$; $N_1$ side) and T·$Au_3$($O_2$; $N_3$ side) (see Figure 2) that exhibit quite similar anchor bonds $Au_7$-$O_2$. We also notice that the difference in the deprotonation energies of the $N_9$-$H_9$ and $N_2$-$H_2'$ groups of G, viz., DPE($N_9$-$H_9$) = 336.4 kcal·mol$^{-1}$ < DPE($N_2$-$H_2'$) = 343.0 kcal·mol$^{-1}$,[13] explains the difference between the hydrogen bonds $N_9$-$H_9$···$Au_{11}$ and $N_2$-$H_2$···$Au_{11}$ in the complexes G·$Au_3$($N_3$; $N_2$ side) and G·$Au_3$($N_3$; $N_9$ side) (see Figure 3).

The changes in the NMR chemical shift $\delta\sigma_{iso}$(H) (prerequisite (vi)) of the bridging proton in the studied N-H···Au H-bonds are all negative and fall within a narrow interval of -1.8 to -3.2 ppm that is close to $\delta\sigma_{iso}$(H) = -2.8 ppm of the water dimer.[11c] This implies that the base - gold interaction induces a deshielding of the bridging proton. The range of the anisotropic shifts $\delta\sigma_{an}$(H) is much wider, from 10.2 to 18.7 ppm, and is also rather close to that of the water dimer.[11c] The largest one, 18.7 ppm, is found for the complex

G·Au$_3$(O$_6$; N$_1$ side).

In general, the data used in prerequisites (iii) – (v) (see also Table 1) provide an estimate of the hydrogen bonding interaction energy, E$_{HB}$. For the nonconventional N-H···Au hydrogen bonding studied here, such an estimation is not possible because of the presence of the anchor bond, that plays a key role in the H-bond formation. We can obtain an approximate upper bound for |E$_{HB}$| by comparing the binding energies, say, of the planar complexes G·Au$_3$(O$_6$; N$_1$ side) and G·Au$_3$(O$_6$; N$_7$ side) (see figure 3). They are characterized by the same Au-O anchor bond, but while G·Au$_3$(O$_6$; N$_1$ side) exhibits a nonconventional N-H···Au hydrogen bond, there is no such bond for G·Au$_3$(O$_6$; N$_7$ side). This gives the upper boun of |E$_{HB}$| ≤ 7.9 kcal·mol$^{-1}$. However, the anchor bond in G·Au$_3$(O$_6$; N$_1$ side) is much stronger than in the G·Au$_3$(O$_6$; N$_7$ side) (2.186 vs. 2.239 Å), so that the upper bound given above should be lowered to 3 - 5 kcal·mol$^{-1}$ that reasonably agrees with the upper-bound found for similar nonconventional hydrogen bondings.[14]

In order to assess basis-set effects, two complexes, A·Au$_3$(N$_3$) and G·Au$_3$(O$_6$; N$_1$ side), respectively with the Au-N and Au-O anchor bonds, were further investigated at the higher computational level B3LYP/RECP (gold) ∪ 6-31++G(d,p) (DNA base). It is shown in Table 1 that the latter produces minor changes in nearly all the examined properties of these two complexes. Two other complexes, A·Au$_4$(N$_3$) and G·Au$_4$(O$_6$; N$_1$ side), with a T-shape four-gold cluster were also studied at this level (Figure 5). Their binding energies amount to 28.8 and 24.2 kcal·mol$^{-1}$, respectively. The former possesses a very short (2.126 Å) Au-N anchor bond. Its N$_9$-H$_9$ bond is directed almost perpendicularly to the centre of the diatomic bond Au$_{11}$-Au$_{12}$ forming two nonconventional H-bonds: N$_9$-H$_9$···Au$_{11}$ and N$_9$-H$_9$···Au$_{13}$, that results in a large Δν(N$_9$-H$_9$) red shift of 275 cm$^{-1}$. In the latter complex, Au-O anchor bond is rather short (2.157

Å), and two almost linear N-H···Au H-bonds are formed with guanine, viz., the $N_1$-$H_1$···$Au_{11}$ with $\Delta\nu(N_1-H_1) = -172$ cm$^{-1}$ and the $N_2$-$H_2$···$Au_{12}$ with $\Delta\nu(N_2-H_2) = -191$ cm$^{-1}$.

In conclusion, we have shown that the bonding between DNA base and odd-size and even-size gold clusters, $Au_3$ and $Au_4$, occurs via the anchor Au-N or Au-O bonds. For all the possible binding sites of DNA, the structural, energetic, and spectroscopic features of the planar and, less stable nonplanar, base-$Au_{3,4}$ complexes have been investigated. A novel type of nonconventional N-H···Au hydrogen bonding, that is formed in the most stable planar complexes between nucleic acid bases and triangle $Au_3$ and T-shape $Au_4$ gold clusters, has been identified. It is the formation of the anchor bond, either Au-N or Au-O, in the planar base-$Au_{3,4}$ complexes that cooperatively, through charge redistribution, "catalyzes" one of the unanchored gold atom to serve as a nonconventional proton acceptor and to form, via its lone pair *5d* orbital, a nonconventional hydrogen bonding with the conventional proton donor of DNA base. In addition, under the assumption of a single-site base-$Au_n$ binding, the reported affinities of DNA bases for $Au_3$ exhibit a fair correlation in magnitude and in relative order with the experimental findings. Note however that some reported anchoring sites of DNA bases are blocked by sugar residues for ssDNA as well as by the intramolecular hydrogen bonds for dsDNA, whereas the remaining sites are available for multi-site bindings. The work on interaction of the DNA duplexes with gold clusters will be published elsewhere [15].

**Acknowledgment.** This work was partially supported by the Région Wallonne (RW. 115012). The computational facilities were provided by NIC (University of Liège) and by F.R.F.C. 9.4545.03 (FNRS, Belgium). E. S. K. gratefully thanks Profs. Lina M. Epstein, Camille Sandorfy, and George V. Yukhnevich for interesting discussions on the N-H···Au hydrogen bonds and valuable suggestions and F.R.F.C. 2.4562.03F for a fellowship.

**References**

(1) (a) Tao, N. J.; de Rose, J. A.; Lindsay, S. M. *J. Phys. Chem.* **1993**, *97*, 910. (b) Mirkin, C. A.; Letsinger, R. L.; Mucic, R. C.; Storhoff, J. J. *Nature* **1996**, *382*, 607. (c) Alivisatos, A. P.; Johnsson, K. P.; Peng, X.; Wislon, T. E.; Loweth, C. J.; Bruchez, M. P., Jr.; Schultz, G. C. *Nature* **1996**, *382*, 609. (d) Storhoff, J. J.; Mirkin, C. A. *Chem. Rev.* **1999**, *99*, 1849. (e) Storhoff, J. J.; Lazarides, A. A.; Mucic, R. C.; Mirkin, C. A.; Letsinger, R. L.; Schatz, G. C. *J. Am. Chem. Soc.* **2000**, *122*, 4640.
(2) (a) Park, S.-J.; Lazarides, A. A.; Mirkin, C. A.; Letsinger, R. L. *Angew. Chem. Int. Ed.* **2001**, *40*, 2909. (b) Niemeyer, C. M. *Angew. Chem. Int. Ed.* **2001**, *40*, 4129. (c) Pirrung, M. C. *Angew. Chem. Int. Ed.* **2002**, *41*, 1277. (d) Harnack, O.; Ford, W. E.; Yasuda, A.; Wessels, J. M. *Nano Lett.* **2002**, *2*, 919.
(3) (a) Parak, W. J.; Pellegrino, T.; Micheel, C. M.; Gerion, D.; Williams, S. C.; Alivisatos, A. P. *Nano Lett.* **2003**, *3*, 33. (b) Alivisatos, A. P. *Nat. Biotechnol.* **2004**, *22*, 47. (c) Daniel, M.-C.; Astruc, D. *Chem. Rev.* **2004**, *104*, 293. (d) Seeman, N. C. *Nature* **2003**, *421*, 427.
(4) (a) Kimura-Suda, H.; Petrovykh, D. Y.; Tarlov, M. J.; Whitman, L. J. *J. Am. Chem. Soc.* **2003**, *125*, 9014. (b) Demers, L. M.; Östblom, M.; Zhang, H.; Jang, N.-H.; Liedberg, B.; Mirkin, C. A. *J. Am. Chem. Soc.* **2002**, *124*, 1128 (c) Storhoff, J. J.; Elghanian, R.; Mirkin, C. A.; Letsinger, R. L. *Langmuir* **2002**, *18*, 6666. (d) Chen, Q.; Frankel, D. J.; Richardson, N. V. *Langmuir* **2002**, *18*, 3219; (e) Giese, B.; McNaughton, D. *J. Phys. Chem. B* **2002**, *125*, 1112. (f) Li, W.; Haiss, W.; Floate, S.; Nichols, R. *Langmuir* **1999**, *15*, 4875. (g) Gourishankar, A.; et al. *J. Am. Chem. Soc.* **2004**, *126*, 13186. (h) Liu, Y.; Meyer-Zaika, W.; Franzka, S.; Schmid, G.; Tsoli, M.; Kuhn, H. *Angew. Chem. Int. Ed.* **2003**, *42*, 2853.
(5) (a) Roelfs, B.; Baumgärtel, H. *Ber. Bunsen-Ges. Phys. Chem.* **1995**, *99*, 677. (b) Xiao, Y. J.; Chen, Y. F. *Spectrochim. Acta A* **1999**, *55*, 1209. (c) Srinivasan, R.; Gopalan, P. *J. Phys. Chem.* **1993**, *97*, 8770. (d) Camargo, A. P. M.; Baumgärtel, H.; Donner, C. *Phys. Chem. Chem. Phys.* **2003**, *5*, 1657 and references therein.
(6) Wells, Jr., D. H.; Delgass, W. N.; Thomson, K. T. *J. Catal.* **2004**, *225*, 69 and references therein.
(7) Ross, R. B.; Powers, J. M.; Atashroo, T.; Ermler, W. C.; LaJohn, L. A.; Christiansen, P. A. *J. Chem. Phys.* **1990**, *93*, 6654.
(8) (a) Remacle, F.; Kryachko, E. S. *Adv. Quantum Chem.* **2004**, *47*, 423. (b) Remacle, F.; Kryachko, E. S. *J. Chem. Phys.* **2005**, *122*, 044304.
(9) Frisch, M. J.; Trucks, G. W.; Schlegel, H. B.; Scuseria, G. E.; Robb, M. A.; Cheeseman, J. R.; Zakrzewski, V. G.; Montgomery, J. A.; Stratmann, R. E.; Burant, J. C.; Dapprich, S.; Millan, J. M.; Daniels, A. D.; Kudin, K. N.; Strain, M. C.; Farkas, O.; Tomasi, J.; Barone, V.; Cossi, M.; Cammi, R.; Mennucci, B.; Pomelli, C.; Adamo, C.; Clifford, S.; Ochterski, J.; Peterson, G. A.; Ayala, P. Y.; Cui, Q.; Morokuma, K.; Malick, D. K.; Rabuck, A. D.; Raghavachari, K.; Foresman, J. B.; Cioslowski, J.; Ortiz, J. V.; Stefanov, B. B.; Liu, G.; Liashenko, A.; Piskorz, P.; Komaromi, I.; Gomperts, R.; Martin, R. L.; Fox, D. J.; Keith, T.; Al-Laham, M. A.; Peng, C. Y.; Nanayakkara, A.; Gonzalez, C.; Challacombe, M.; Gill, P. M. W.; Johnson, B. G.; Chen, W.; Wong, M. W.; Andres, J. L.; Head-Gordon, M.; Replogle, E. S.; Pople, J. A. *GAUSSIAN 03 (Revision A.1)*, Gaussian, Inc., Pittsburgh, PA, **2003**.
(10) (a) Pimentel, C. G.; McClellan, A. L. *The Hydrogen Bond*; W. H. Freeman: San

Francisco, **1960**. (b) *The Hydrogen Bond. Recent Developments in Theory and Experiments*, Schuster, P.; Zundel, G.; Sandorfy, C., Eds.; North-Holland: Amsterdam, **1976**. (c) Jeffrey, G. A.; Saenger, W. *Hydrogen Bonding in Biological Structures*; Springer: Berlin, **1991**. (d) Scheiner, S. *Hydrogen Bonding. A Theoretical Perspective*; Oxford University Press: Oxford, **1997**. (e) Desiraju, G. R.; Steiner, T. *The Weak Hydrogen Bond in Structural Chemistry and Biology*; Oxford University Press: Oxford, **1999**. (f) Steiner, T. *Angew. Chem. Int. Ed*. **2002**, *41*, 48.

(11) (a) Hinton, J. F.; Wolinski, K. in *Theoretical Treatments of Hydrogen Bonding,* Hadži, D., Ed.; Wiley: Chichester, **1997**; p. 75. (b) Becker, E. D. in *Encyclopedia of Nuclear Magnetic Resonance*, Grant, D. M.; Harris, R. K., Eds.; Wiley: New York, **1996**; p. 2409. (c) Kar, T.; Scheiner, S. *J. Phys. Chem. A* **2004**, *108*, 9161 and references therein.

(12) (a) Epstein, L. M.; E. S. Shubina, *Coord. Chem. Rev*. **2002**, *231*, 165. (b) Brammer, L. *Dalton Trans*. **2003**, 3145 and references therein.

(13) (a) Chandra, A. K.; Nguyen, M. T.; Uchimaru, T.; Zeegers-Huyskens, T. *J. Phys. Chem. A* **1999**, *103*, 8853. (b) Kryachko, E. S.; Nguyen, M. T.; Zeegers-Huyskens, T. *J. Phys. Chem. A* **2001**, *105*, 1288, 1934.

(14) Kryachko, E. S.; Remacle, F. *Chem. Phys. Lett*. **2005**, *404*, 142.

(15)  Kryachko, E. S.; Remacle, F. *J. Phys. Chem. B* (submitted).

***Table 1.*** Some key features of the planar DNA base-$Au_3$ complexes with the N-H···Au H-bond. Three H···Au bond lengths, $r(H_3···Au_8)$ = 2.883 Å in T·$Au_3(O_4)$, $r(H_2'···Au_{11})$ = 2.890 Å in G·$Au_3(N_3; N_2$ side), and $r(H_3···Au_8)$ = 2.913 Å in T·$Au_3(O_2; N_3$ side) slightly exceed the sum of van der Waals radii (see also[10f] and comment[21] therein about an extension of the van der Waals cutoff to 3.0 – 3.2 Å). The binding energy $E_b$ (including zero-point vibrational energy) and the enthalpy of formation, $-\Delta H_f$, are given in kcal·mol$^{-1}$ and defined with respect to the infinitely separated base and $Au_3$ cluster. $\Delta\nu$(N-H) is in cm$^{-1}$. $R_{IR}$ is the ratio of the IR activities of the corresponding N-H stretches in the H-bonds and in the bases. $\delta\sigma_{iso}$ and $\delta\sigma_{an}$ are the NMR shifts (in ppm) taken with respect to the corresponding monomers. The data in parentheses refer to the level B3LYP/RECP (gold) ∪ 6-31++G(d,p) (DNA base). Bond lengths are given in Å and angles in deg.

| Complex | $E_b$ | $-\Delta H_f$ | Anchor bond | $\Delta R$(N-H) | r(H···Au) | ∠N-H···Au | $\Delta\nu$(N-H) | $R_{IR}$ | $\delta\sigma_{iso}$ | $\delta\sigma_{an}$ |
|---|---|---|---|---|---|---|---|---|---|---|
| A·$Au_3(N_1)$ | 22.6 | 22.7 | 2.153 | 0.009 | 2.836 | 175.2 | 153 | 5.6 | -2.4 | 10.3 |
| A·$Au_3(N_3)$ | 24.4 | 24.0 | 2.138 | 0.014 | 2.698 | 160.8 | 252 | 8.7 | -2.4 | 13.0 |
|  | (24.0) |  |  | (0.007) |  |  | (270) |  |  |  |
| A·$Au_3(N_7)$ | 22.3 | 21.9 | 2.130 | 0.007 | 2.816 | 165.1 | 116 | 9.0 | -2.2 | 14.0 |
| T·$Au_3(O_2; N_1$ side) | 14.4 | 13.9 | 2.218 | 0.017 | 2.608 | 178.8 | 324 | 11.0 | -2.9 | 16.6 |
| T·$Au_3(O_2; N_3$ side) | 10.8 | 10.3 | 2.227 | 0.011 | 2.913 | 171.8 | 199 | 9.0 | -1.9 | 13.9 |
| T·$Au_3(O_4)$ | 12.4 | 11.9 | 2.209 | 0.013 | 2.883 | 174.4 | 224 | 9.0 | -2.2 | 14.1 |
| G·$Au_3(N_3; N_2$ side) | 20.7 | 20.1 | 2.147 | 0.009 | 2.890 | 176.1 | 115 | 9.0 | -2.5 | 10.2 |
| G·$Au_3(N_3; N_9$ side) | 20.9 | 20.3 | 2.146 | 0.010 | 2.841 | 161.8 | 181 | 6.0 | -1.8 | 11.7 |
| G·$Au_3(O_6; N_1$ side) | 18.4 | 17.9 | 2.186 | 0.015 | 2.580 | 173.1 | 302 | 15.0 | -3.2 | 18.7 |
|  | (18.4) |  |  | (0.012) |  |  | (324) |  |  |  |
| C·$Au_3(O_2; N_1$ side) | 20.0 | 19.5 | 2.177 | 0.016 | 2.627 | 178.9 | 306 | 14.0 | -3.2 | 17.6 |
| C·$Au_3(N_3)$ | 25.4 | 25.1 | 2.164 | 0.014 | 2.673 | 179.7 | 232 | 8.0 | -3.2 | 12.6 |

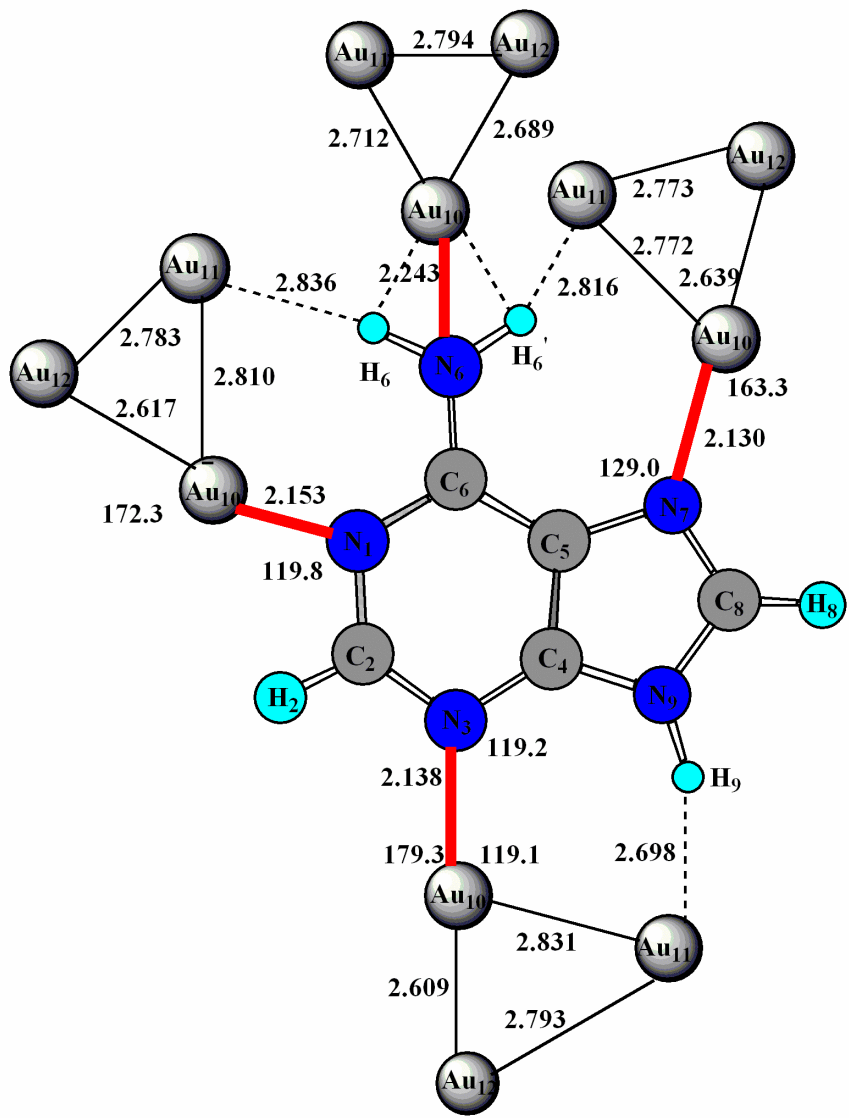

*Figure 1*. The four possible planar ($N_1$, $N_3$, $N_7$) and nonplanar ($N_6$) binding sites of the gold cluster $Au_3$ to adenine. Also shown is the $NH_2$ anchored complex $A·Au_3$ ($N_6$). For each complex, the anchor bond is drawn as a thick (red) line and the nonconventional H-bond as a dotted line. The stability ordering of the complexes is (see also Table 1): $A·Au_3(N_3) > A·Au_3(N_1) > A·Au_3(N_7) > A·Au_3(N_6)$. The bond lengths are given in Å and bond angles in deg.

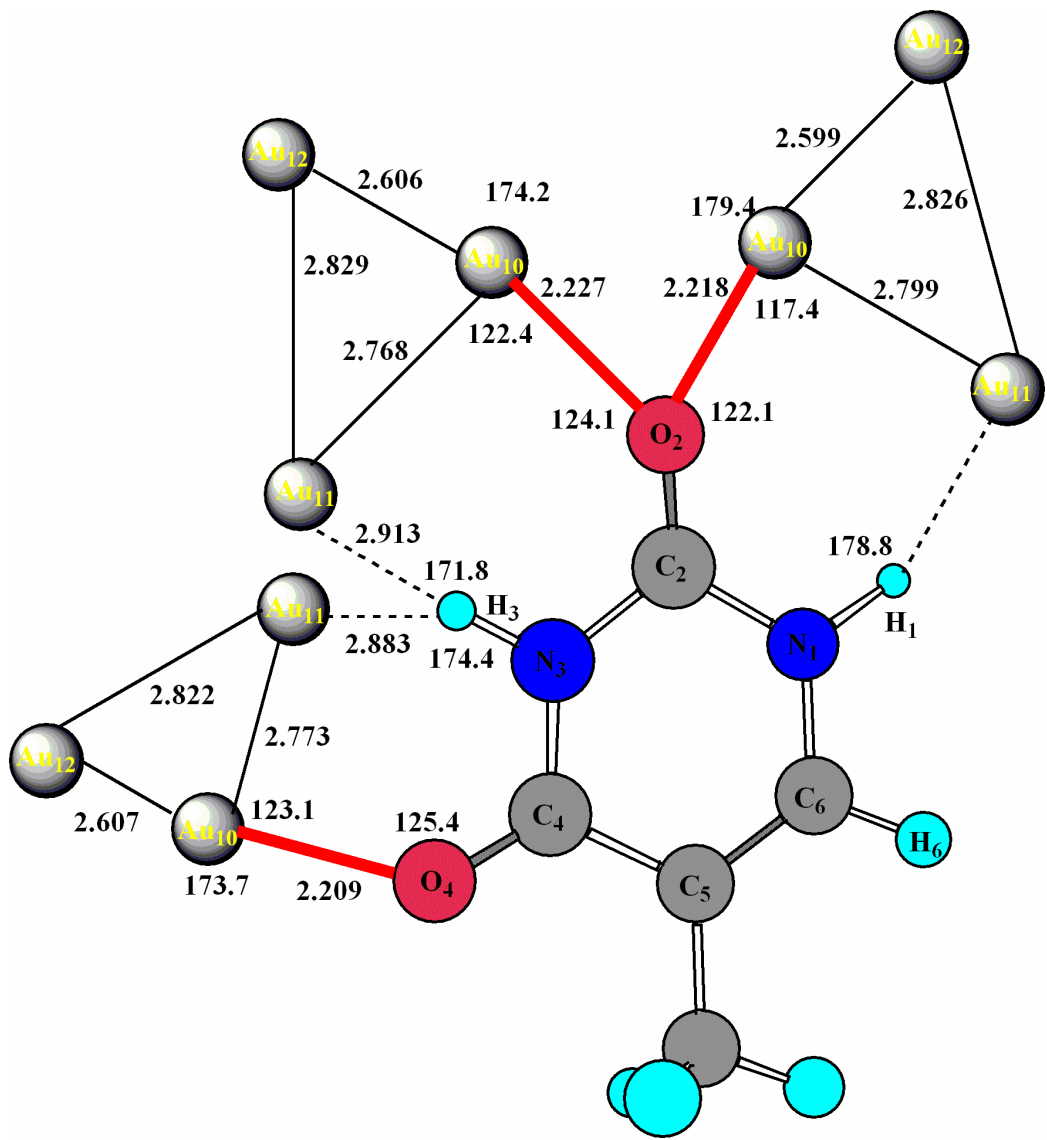

*Figure 2*. The three possible planar (O$_2$(N$_1$), O$_2$(N$_3$), O$_4$) binding sites of the gold cluster Au$_3$ to thymine. For each complex, the anchor bond is drawn as a thick (red) line and the nonconventional H-bond as a dotted line. The stability ordering of the complexes is (see also Table 1): T·Au$_3$(O$_2$; N$_1$ side) > T·Au$_3$(O$_4$) > T·Au$_3$(O$_2$; N$_3$ side). The bond lengths are given in Å and bond angles in deg.

**Figure 3.** The six possible planar ($N_3(N_2)$, $N_3(N_9)$, $O_6(N_1)$, $O_6(N_7)$, $N_7$) and nonplanar($N_2$) binding sites of the gold cluster $Au_3$ to guanine. For each complex, the anchor bond is drawn as a thick (red) line and the nonconventional H-bond as a dotted line. The anchoring in $N_2$ is to the amino group. The stability ordering of the complexes is (see also Table 1): $G \cdot Au_3(N_3; N_9 \text{ side}) > G \cdot Au_3(N_3; N_2 \text{ side}) > G \cdot Au_3(N_7) > G \cdot Au_3(O_6; N_1 \text{ side}) > G \cdot Au_3(O_6; N_7 \text{ side}) > G \cdot Au_3(N_2)$. The bond lengths are given in Å and bond angles in deg.

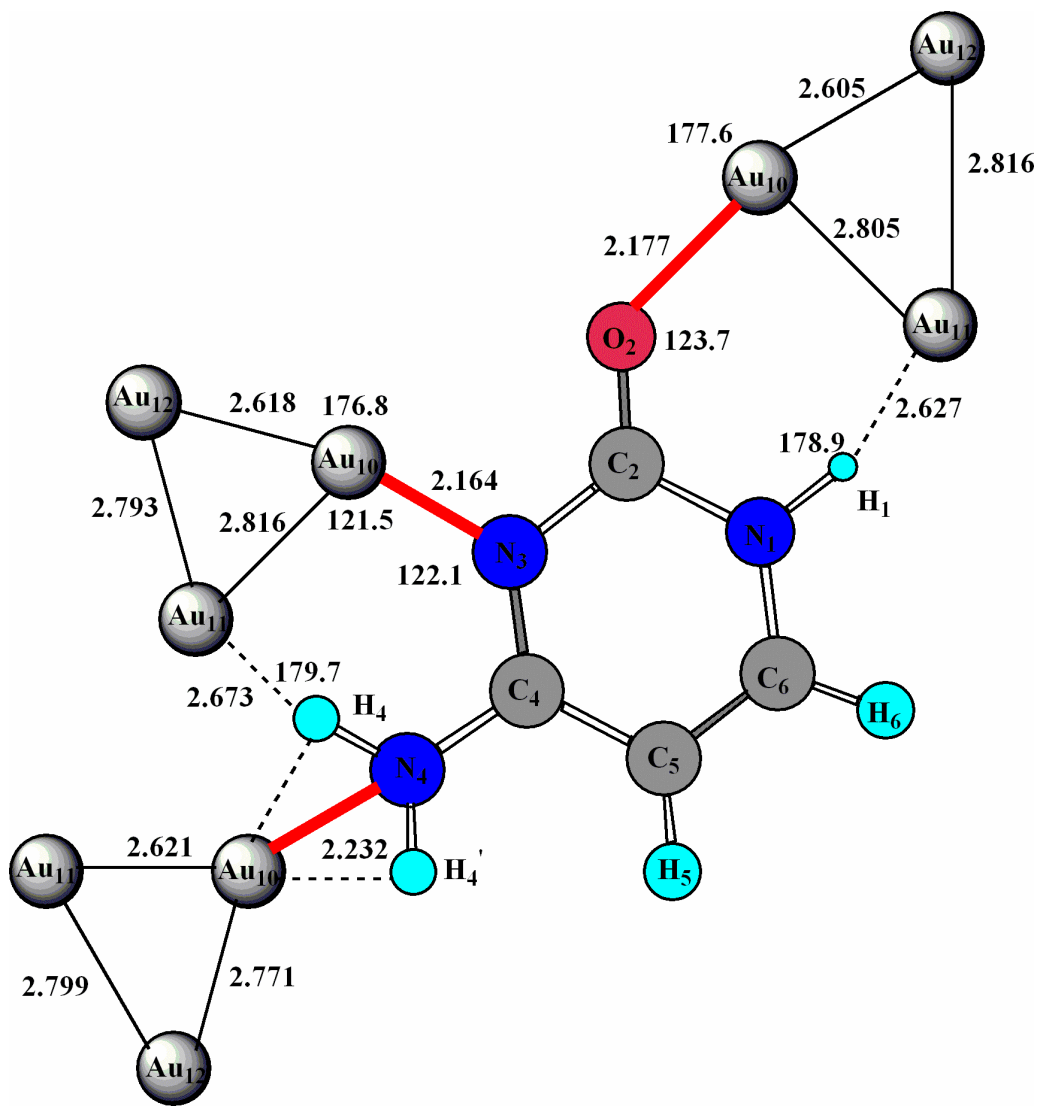

*Figure 4*. The three possible planar ($O_2(N_1)$, $N_3$) and nonplanar ($N_4$) binding sites of the gold cluster $Au_3$ to cytosine. For each complex, the anchor bond is drawn as a thick (red) line and the nonconventional H-bonds in dotted lines. For the binding site $N_4$, the anchor bond is to the $NH_2$ group. The stability ordering of the complexes is (see also Table 1): $C \cdot Au_3(N_3) > C \cdot Au_3(O_2; N_1 \text{ side}) > C \cdot Au_3(N_4)$. The bond lengths are given in Å and bond angles in deg.

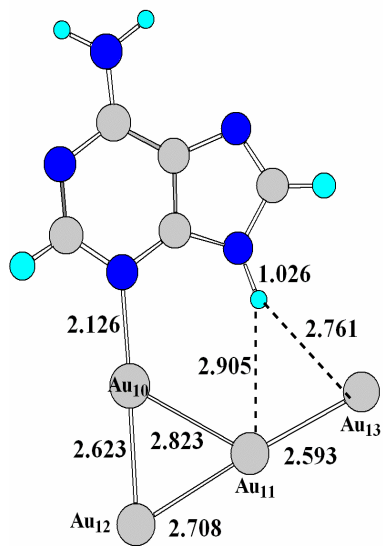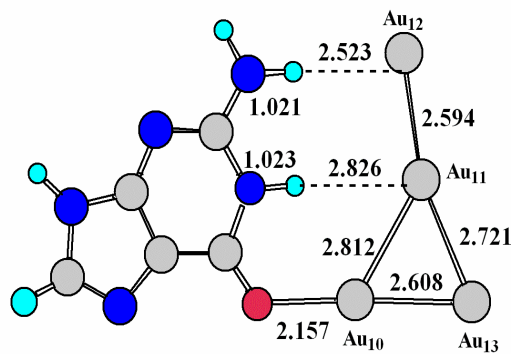

***Figure 5.*** Complexes A·Au$_4$(N$_3$) and G-Au$_4$(O$_6$; N$_1$ side) with a T-shape four-gold cluster. The bond lengths are given in Å and bond angles in deg. Atomic numbering is indicated in Figures 1 and 3. The H-bonds are drawn in dotted lines.

Table of content graphical abstract

## Complexes of DNA Bases and Gold Clusters $Au_3$ and $Au_4$ Involving Nonconventional N-H...Au Hydrogen Bonding

**E. S. Kryachko and F. Remacle**

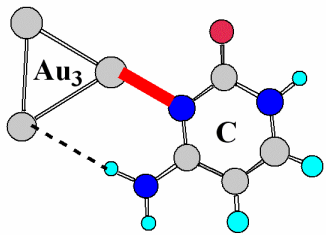

Legend : The most stable complex formed between the gold cluster $Au_3$ and cytosine. The complex is formed via an anchor N-Au bond (thick red line), that is reinforced by a nonconventional N-H···Au hydrogen bond (dotted line).